\documentclass[aps,prd,twocolumn,groupedaddress,showpacs,nofootinbib,amssymb]{revtex4-1}
\usepackage[T1]{fontenc}
\usepackage[latin1]{inputenc}
\usepackage{graphicx}
\usepackage[english]{babel}
\usepackage{graphicx}
\usepackage{bm}
\usepackage{amsmath}
\usepackage{amssymb}
\usepackage{amsfonts}
\usepackage{epsfig}
\usepackage{colordvi}
\usepackage{color}

\begin{document}
\def\ben{\begin{eqnarray}}
\def\een{\end{eqnarray}}
\def\be{\begin{equation}}
\def\ee{\end{equation}}
\def\ba{\begin{eqnarray}}
\def\ea{\end{eqnarray}}
\def\bea{\begin{eqnarray*}}
\def\eea{\end{eqnarray*}}
\def\D{\partial}
\def\na{\nabla}
\def\ie{{\it i.e.} }
\def\eg{{\it e.g.} }
\def\etc{{\it etc.} }
\def\etal{{\it et al.}}
\def\nn{\nonumber}
\def\ra{\rightarrow}
\def\lra{\leftrightarrow}
\def\lsim{\buildrel{<}\over{\sim}}
\def\gsim{\buildrel{>}\over{\sim}}
\def\G{\Gamma}

\title{Gravitational Cherenkov Radiation from  Extended Theories of Gravity}

\author{M. De Laurentis$^{1,2}$,   S. Capozziello$^{1,2}$, G. Basini$^{3}$}
\affiliation{ \it 
 $^1$Dipartimento di Scienze Fisiche, Università
di Napoli {}``Federico II'', Compl. Univ. di
Monte S. Angelo, Edificio G, Via Cinthia, I-80126, Napoli, Italy\\
$^2$INFN Sezione  di Napoli, Compl. Univ. di
Monte S. Angelo, Edificio G, Via Cinthia, I-80126, Napoli, Italy,\\
$^3$ Laboratori Nazionali di Frascati, INFN, Via E. Fermi,  C.P. 13, I-0044 Frascati, Italy.}
\date{\today}

\begin{abstract}
We linearize the field equations for higher order theories of gravity that contain scalar invariants other than
the Ricci scalar. We find that besides a massless spin-2 field (the standard graviton), the theory
contains also spin-0 and spin-2 massive modes with the latter being, in general, ghost modes.
The rate at which such particles
would emit gravitational Cherenkov radiation is calculated for some interesting physical cases.
\end{abstract}

\maketitle


Issues coming from Cosmology and
Quantum Field Theory suggest to extend  General Relativity
(GR) in order to cure several shortcomings emerging from
astrophysical observations and fundamental physics. For example,
problems in the early time cosmology led to the conclusion that the Standard Cosmological Model could be inadequate to describe
the Universe at extreme regimes. In fact, GR does not work at the
fundamental level, when one wants to achieve a full quantum
description of space-time (and then of gravity).
On the other hand, Einstein gravity seems to present problems also at infrared scales since the so called dark sector (dark energy and dark matter) could be cured either detecting new fundamental particles constituting the largest part of cosmic matter-energy or claiming for alternative or extended approach to the gravitational interaction.
Given these facts and the lack of a final self-consistent Quantum
Gravity Theory, Extended Theories of Gravity (ETGs) have been pursued
as part of a semi-classical scheme where GR and its positive results
should be recovered \cite{PRnostro}. The approach  is based on corrections and enlargements of the Einstein theory
adding higher-order curvature invariants and minimally or non-minimally coupled
scalar fields into dynamics which come out from  effective actions of unified theories formulated on curved space-time \cite{birrell}.

Beside fundamental physics motivations,
these theories have received a lot of interest in cosmology since
they naturally exhibit inflationary behavior which can overcome
the shortcomings of standard cosmology. The related cosmological
models seem realistic and capable of coping with observations.
ETGs  play an interesting role also in  describing the today observed
universe. In fact, the good quality data of last decade has
made it possible to shed new light on the cosmic effective picture 

From an astrophysical point of view, ETGs do not require finding
candidates for dark energy and dark matter at the fundamental
level; the approach starts from taking into account only the
observed ingredients (i.e. gravity, radiation and baryonic matter);
it is in full agreement with the early spirit of a GR that could
not act in the same way at all scales \cite{francaviglia1,francaviglia2,francaviglia3,francaviglia4,francaviglia5}. 

At a fundamental level, the main features of ETGs is the emergence of new gravitational modes that can be roughly classified as tensor and scalar modes related to massive, massless and ghost particles \cite{greci}.
In this view,  GR is an exception including only massless tensor modes. 
 The further modes  could have interesting  effects at ultra-violet and infra-red scales and could play an important role both in  the Standard Model of Particles  and in gravitational radiation \cite{felix}. In particular, they could  be connected to the symmetry breaking  in the  high energy limit (and  investigated at LHC \cite{LHC}) and have a signature in  the cosmological  stochastic background of gravitational waves \cite{basini,greci,francaviglia2}.

In this paper we want  to investigate possible  Cherenkov-like  radiation effects \cite{gupta,gaetano,mpla} related to these further gravitational modes that could have observational effects.
For our aim,
assuming  any  curvature invariant other than the Ricci scalar \footnote{ We restrict  to fourth-order theories which  give the main contributions in  renormalization process \cite{birrell}, but  the following considerations can be extended to any higher-order model involving generic powers of the $\Box$-operator and combinations of curvature invariants.}
 a generic  action for the gravitational interaction is \cite{PRnostro}
 \be {\cal A}=\int
d^4x\sqrt{-g} f(R,P,Q)\,. \ee where \ba && P\equiv R_{\alpha\beta}R^{\alpha\beta}\nn \\
&&Q\equiv R_{\alpha\beta\gamma\delta}R^{\alpha\beta\gamma\delta}\,. \ea
Varying with respect to the metric, one gets the field equations:
 \ba FG_{\mu\nu}&=&\frac{1}{2}g_{\mu\nu}\left(f-
R~F\right)-(g_{\mu\nu}\Box-\na_\mu\na_\nu)F\nn\\&&
-2\left(f_P R^\alpha_\mu
R_{\alpha\nu}+f_Q~R_{\alpha\beta\gamma\mu}R^{\alpha\beta\gamma}_{~~~\nu} \right)\nn\\&&-
g_{\mu\nu}\na_\alpha\na_\beta(f_P R^{\alpha\beta})-\Box (f_P
R_{\mu\nu})\nn\\ &&+2\na_\alpha\na_\beta\left(f_P~R^\alpha_{~(\mu}\delta^beta_{~\nu)}+2
f_Q~R^{\alpha~~~~\beta}_{~(\mu\nu)}\right)\,,\nn\\\label{fieldeqs}\ea
where we have set
  \be F\equiv\frac{\D f}{\D R}, ~~~f_P\equiv\frac{\D f}{\D P}, ~~~f_Q\equiv\frac{\D
f}{\D Q} \,,\ee and  $\Box=g^{\alpha\beta}\na_\alpha\na_\beta$ is the d'Alembert
operator \footnote{The notation $T_{(\mu\nu)}=\frac{1}{2}(T_{\mu\nu}+T_{\nu\mu})$
denotes symmetrization with respect to the indices $(\mu,\nu)$.}.

Taking the trace of Eq. (\ref{fieldeqs}),  we find:
 \ba && \Box\left(F+\frac{f_P}{3}
R\right)=\nonumber\\&&
\frac{1}{3}\left[ 2 f-RF-2 \na_a\na_b((f_P+2f_Q)R^{\alpha\beta})+\right.\nonumber\\
&&\left.-2
(f_P P+f_Q Q)\right] \label{trace}\,.\ea
Expanding the third term on the r.h.s. of (\ref{trace}) and using the
Bianchi identity $G^{\alpha\beta}_{~~;\beta}=0$, we get: \ba
&&\Box\left(F+\frac{2}{3}(f_P+f_Q) R\right)=
\frac{1}{3}\times\nn\\ && [2
f-RF-2R^{\alpha\beta}\na_\alpha\na_\beta(f_P+2f_Q)-R\Box(f_P+2f_Q)\nn\\ &&-2 (f_P
P+f_Q Q)] \label{trace1}\,.\ea
If we define \ba \phi & \equiv&
F+\frac{2}{3}(f_P+f_Q) R\,, \label{phidef} 
\ea
and
\ba
\frac{dV}{d\phi} & \equiv& \textrm{r.h.s. }\nn \ea
of (\ref{trace1}),
 we get a Klein-Gordon equation for the scalar field $\phi$:
\be \Box \phi = \frac{dV}{d\phi}\,. \ee
It is  clear that the scalar field $\phi$, 
assumes the role of a field induced by the further degrees of freedom of ETGs.
Obviously, $\phi$ is identically zero in GR.

Let us linearize around the Minkowski background and then
we assume

\ba g_{\mu\nu}&=&\eta_{\mu\nu}+h_{\mu\nu}\,, \nn \\\\
\phi&=&\phi_0+\delta \phi\,. \ea

Then, from Eq. (\ref{phidef}), we get
\be \delta \phi=\delta F+\frac{2}{3}(\delta f_P+\delta f_Q) R_0+
\frac{2}{3}(f_{P0}+f_{Q0}) \delta R \label{pertphi1}\,,\ee
where $R_0
\equiv R(\eta_{\mu\nu})=0$ and similarly $f_{P0}=\frac{\D f}{\D
P}|_{\eta_{\mu\nu}}$  which is either constant or zero.  Note that the index  $0$ indicates quantities calculated  with respect to
the Minkowski metric, that means no deformation.  $\delta
R$  denotes the first order perturbation of the Ricci scalar
which, along with the perturbed parts of the Riemann and Ricci
tensors, are given by:

\ba \delta R_{\mu\nu\rho\sigma}&=&\frac{1}{2}\left(\D_\rho \D_\nu
h_{\mu \sigma}+\D_\sigma \D_\mu h_{\nu \rho}-\D_\sigma \D_\nu
h_{\mu
\rho}-\D_\rho \D_\mu h_{\nu \sigma} \right)\,, \nn\\
\delta R_{\mu\nu} &=& \frac{1}{2}\left(\D_\sigma \D_\nu
h^\sigma_{~\mu}+\D_\sigma \D_\mu h^\sigma_{~\nu}-\D_\mu \D_\nu
h-\Box h_{\mu \nu} \right)\,,\nn\\
\delta R &=& \D_\mu \D_\nu h^{\mu \nu}-\Box h\,,\nn\ea where
$h=\eta^{\mu \nu} h_{\mu \nu}$. The first term of Eq.
(\ref{pertphi1}) is \be \delta F=\frac{\D F}{\D R}|_0~\delta
R+\frac{\D F}{\D P}|_0~\delta P+\frac{\D F}{\D Q}|_0~\delta Q\,. \ee
However, since $\delta P$ and $\delta Q$ are second order, we get
$\delta F\simeq F_{,R0}~ \delta R $ and
 \be
 \delta \Phi
=\left(F_{,R0} +\frac{2}{3} (f_{P0}+f_{Q0})\right) \delta R\,.
\label{pertphi2}
\ee
Finally, from Eq. (\ref{trace1}) we get the
Klein-Gordon equation for the scalar perturbation $\delta \phi$
\ba \Box \delta \phi&=&\frac{1}{3}\frac{F_0}{F_{,R0} +\frac{2}{3}
(f_{P0}+f_{Q0})}\delta \phi-\nn\\ &&+\frac{2}{3}\delta
{R}^{\alpha\beta}\D_\alpha\D_\beta(f_{P0}+2f_{Q0})-\frac{1}{3}\delta
{R}\Box(f_{P0}+2f_{Q0})\nn \\ &=& m_s^2 \delta \phi
\label{kgordon1}\,.\nn\\ \ea The last two terms in the first line are
actually  zero since the terms $f_{P0}$, $f_{Q0}$ are constants
and we have defined the scalar mass as $m_s^2\equiv
\frac{1}{3}\frac{F_0}{F_{,R0} +\frac{2}{3} (f_{P0}+f_{Q0})}$.

Perturbing the field equations (\ref{fieldeqs}) we get: \ba &&
F_0(\delta{R}_{\mu\nu}-\frac{1}{2}\eta_{\mu\nu}
\delta{R})=\nn\\ &&-(\eta_{\mu\nu}\Box -\D_\mu\D_\nu)(\delta
\phi-\frac{2}{3}(f_{P0}+f_{Q0})\delta{R})\nn
\\&&-\eta_{\mu\nu} \D_\alpha\D_\beta (f_{P0} \delta{R}^{\alpha\beta})-\Box(f_{P0}
\delta{R}_{\mu\nu})\nn\\ &&+2
\D_\alpha\D_\beta(f_{P0}~\delta{R}^\alpha_{~(\mu}\delta^\beta_{~\nu)}+2
f_{Q0}~\delta{R}^{\alpha~~~~\beta}_{~(\mu\nu)})\,.\nn\\ \ea
It is convenient to
work in Fourier space where the following substitutions have to be operated:  $\D_\gamma
h_{\mu\nu}\rightarrow i k_\gamma h_{\mu\nu}$ and $\Box h_{\mu\nu}
\rightarrow -k^2 h_{\mu\nu}$.
The above equation becomes
\ba
&& F_0(\delta{R}_{\mu\nu}-\frac{1}{2}\eta_{\mu\nu}
\delta{R})=\nn\\ &&(\eta_{\mu\nu}k^2 -k_\mu k_\nu)(\delta
\phi-\frac{2}{3}(f_{P0}+f_{Q0})\delta{R})\nn
\\&&+\eta_{\mu\nu} k_\alpha k_\beta (f_{P0} \delta{R}^{\alpha\beta})+k^2(f_{P0}
\delta{R}_{\mu\nu})\nn\\ &&-2 k_a
k_b(f_{P0}~\delta{R}^a_{~(\mu}\delta^b_{~\nu)})-4 k_\alpha k_\beta(
f_{Q0}~\delta{R}^{\alpha~~~~\beta}_{~(\mu\nu)})\,.\nn\\ \label{fields2}\ea
We
can rewrite the metric perturbation as \be
h_{\mu\nu}=\bar{h}_{\mu\nu}-\frac{\bar{h}}{2}~
\eta_{\mu\nu}+\eta_{\mu\nu} h_f \,,\label{gauge}\ee and impose  the standard gauge conditions
$\D_\mu
\bar{h}^{\mu\nu} =0$ and $\bar{h}=0$. The first of these
conditions implies that $k_\mu \bar{h}^{\mu\nu} =0$ while the
second gives \ba h_{\mu\nu}&=&\bar{h}_{\mu\nu}+\eta_{\mu\nu} h_f\,, \nn \\
h&=&4 h_f\,.\ea
Inserting into the perturbed curvature quantities, we get
 \ba \delta
R_{\mu\nu}&=&\frac{1}{2}\left(2k_\mu k_\nu h_f+k^2 \eta _{\mu\nu}
h_f+k^2 \bar{h}_{\mu\nu}\right)\,, \nn\\
\delta R &=& 3k^2 h_f\,,\nn\\
k_\alpha k_\beta ~\delta
R^{\alpha~~~~~\beta}_{~~(\mu\nu)~}&=&-\frac{1}{2}\left((k^4
\eta_{\mu\nu}-k^2 k_\mu k_\nu)h_f+k^4 \bar{h}_{\mu\nu}\right)\,,\nn\\
k_\alpha k_\beta~\delta{R}^\alpha_{~(\mu}\delta^\beta_{~\nu)}&=&\frac{3}{2}k^2k_\mu
k_\nu h_f\,. \nn\\ \label{results1}\ea
Substituting  Eqs.
(\ref{gauge})-(\ref{results1}) into (\ref{fields2}) and after some
algebra we get: \ba
&&\frac{1}{2}\left(k^2-k^4\frac{f_{P0}+4f_{Q0}}{F_0}\right)\bar{h}_{\mu\nu}=\nn\\
&&(\eta_{\mu\nu}k^2 -k_\mu k_\nu)\frac{\delta \phi}{F_0}
+(\eta_{\mu\nu}k^2 -k_\mu k_\nu)h_f\,, \nn\\\ea
Defining $h_f\equiv
-\frac{\delta \phi}{F_0}$ we find the equation for the
perturbations:
\be
\left(k^2+\frac{k^4}{m^2_{spin2}}\right)\bar{h}_{\mu\nu}=0\,,
\label{solution} \ee
 where  $m^2_{spin2}\equiv
-\frac{F_0}{f_{P0}+4f_{Q0}}$.
 From Eq. (\ref{kgordon1}) we
get: \be \Box h_f=m_s^2 h_f \,.\label{kgordon3}\ee
From Eq.
(\ref{solution}) it is easy to see that we have a modified
dispersion relation which corresponds to a massless spin-2 field
($k^2=0$) and a massive spin-2  field
$k^2=\frac{F_0}{\frac{1}{2}f_{P0}+2f_{Q0}}\equiv -m^2_{spin2}$.
 To see  better this point, let us  note that the propagator for
$\bar{h}_{\mu\nu}$ can be rewritten as \be G(k) \propto
\frac{1}{k^2}-\frac{1}{k^2+m^2_{spin2}}\,. \ee Clearly the second
term has the opposite sign, which indicates the presence of a
ghost energy mode (see also
\cite{Nunez:2004ts,Chiba:2005nz,Stelle:1977ry}).

As a "sanity check", we can see that for the Gauss-Bonnet term
$\mathcal{L}_{GB}=Q-4P+R^2$ we have $f_{P0}=-4$ and $f_{Q0}=1$.
Then, Eq. (\ref{solution}) simplifies to $k^2
\bar{h}_{\mu\nu}=0$ and, in this case, we have no negative
energy modes as expected.

The solutions of Eqs. (\ref{solution}) and (\ref{kgordon3}) can be
written in terms of plane waves \ba \bar{h}_{\mu\nu}&=&e_{\mu\nu}
(\overrightarrow{p}) \cdot  exp(ik^\alpha x_\alpha)+c.c. \label{pw1}
\ea \ba h_f &=& e(\overrightarrow{p}) \cdot exp(iq^\alpha
x_\alpha)+c.c.\label{pw2} \ea where

\begin{equation}
\begin{array}{ccc}
k^{\alpha}\equiv(\omega_{m_{spin2}},\overrightarrow{p}) &  & \omega_{m_{spin2}}=\sqrt{m_{spin2}^{2}+p^{2}}\\
\\q^{\alpha}\equiv(\omega_{m_s},\overrightarrow{p}) &  & \omega_{m_s}=\sqrt{m_s^{2}+p^{2}}.\end{array}\label{eq: k e q}\end{equation} and
where $m_{spin2}$ is zero (non-zero) in the case of massless
(massive) spin-2 mode and the polarization tensors $e_{\mu\nu}
(\overrightarrow{p})$ can be found in Ref. \cite{vanDam:1970vg}
(see equations (21)-(23)).  Eqs. (\ref{solution}) and
(\ref{pw1}), contain the equation and the solution for the standard gravitational waves
of GR \cite{gravitation} plus massive spin 2 terms\footnote{Here $+c.c.$ means plus the complex conjugate of the preceding term. }.
Eqs. (\ref{kgordon3}) and (\ref{pw2}) are respectively the
equation and the solution for the massive scalar mode (see also
\cite{felix}).

The fact that the dispersion law for the modes of the massive
field $h_{f}$ is not linear has to be emphasized. The velocity of
every {} ordinary (arising from GR)
mode $\bar{h}_{\mu\nu}$ is the light speed $c$, but the dispersion
law (the second of Eq. (\ref{eq: k e q})) for the modes of $h_{f}$
is that of a massive field which can be seen as a
wave-packet. Also, the group-velocity of a
wave-packet of $h_{f}$, centered in $\overrightarrow{p}$, is

\begin{equation}
\overrightarrow{v_{G}}=\frac{\overrightarrow{p}}{\omega},\label{eq: velocita' di gruppo}\end{equation}
which is exactly the velocity of a massive particle with mass $m$
and momentum $\overrightarrow{p}$.
From the second of Eqs. (\ref{eq: k e q}) and Eq. (\ref{eq: velocita' di gruppo})
it is straightforward to obtain:

\begin{equation}
v_{G}=\frac{\sqrt{\omega^{2}-m^{2}}}{\omega}.\label{eq: velocita' di gruppo 2}\end{equation}
This means that the  speed of the wave-packet is
\begin{equation}
m=\sqrt{(1-v_{G}^{2})}\omega.\label{eq: relazione massa-frequenza}
\end{equation}
Summarizing these results, we can say that considering ETGs (which we have generically  assumed as  analytic functions of curvature invariants)  more gravitational modes than the standard massless ones of GR have to be taken into account. 
 In fact, we can note that there are two
conditions for Eq. (\ref{kgordon1}) that depend on the value of
$k^2$. In fact  we  have a $k^2=0$ mode that corresponds to a
massless spin-2 field with two independent polarizations plus a
scalar mode, while if we have $k^2\neq0$ we have a massive spin-2
ghost mode and there are five independent polarization tensors
plus a scalar mode \cite{greci}. 
In particular taking $\overrightarrow{p}$ in the $z$ direction, 
 we have 6 polarizations \footnote{The polarizations are
defined in our 3-space, not in a spacetime with extra
dimensions. Each polarization mode is orthogonal to 
another and it is normalized $e_{\mu\nu}e^{\mu\nu} =2\delta$. Note that these further modes are not traceless, in contrast to the ordinary
plus and cross polarization modes of GR.} 
and the amplitude can be written as

\begin{eqnarray}
h_{\mu\nu}&=&\left[e_{\mu\nu}^{(+)}+e_{\mu\nu}^{(\times)}
+e_{\mu\nu}^{(B)}+e_{\mu\nu}^{(C)}+\right.
\nonumber\\ &&+\left.e_{\mu\nu}^{(D)}+e_{\mu\nu}^{s}\right]\cdot  exp(ik^\alpha x_\alpha)+c.c \label{hpol}\end{eqnarray} 
where the first two terms describe the standard polarizations of gravitational waves  arising from GR, while the others are the 
massive fields arising from  higher order theories.
At this point we can calculate the gravitational radiation. To  this end,  we have to calculate the average energy momentum tensor of a plane waves, that, in terms of polarization tensors, can be expressed as

\begin{eqnarray}
\left\langle T_{\mu\nu}\right\rangle=\frac{k_\mu k_\nu}{16\pi G}\left(e^{\lambda\rho*}e_{\lambda\rho}-\frac{1}{2}|e^{\lambda}\,_{\lambda}|^2\right)\,,\label{tensopol}
\end{eqnarray}
where we have denoted with '$*$' the complex conjugate and the brackets $\langle\cdot\cdot\rangle$ indicate the average process.
Note that the formula (\ref{tensopol}) have to be calculated for each polarization.

For our aim, it is  convenient to write the polarization tensor  $e_{\mu\nu}$ in Eq. (\ref{pw1}) explicitly in terms of the Fourier transform of energy momentum tensor $T_{\mu\nu}$ of a system that emits gravitational radiation \footnote{Note that here, $e_{\mu\nu}$, is the sum of the polarization tensors arising from higher order gravity, as shown in Eq.(\ref{hpol}).} \cite{quadru}:
\begin{equation}
e_{\mu\nu}({\bf x},\omega)=\frac{4G}{r}\left[T_{\mu\nu}({\bf k},\omega)-\frac{1}{2}\eta_{\mu\nu}T^{\lambda}\,_{\lambda}({\bf k},\omega)\right]\,,
\label{A}
\end{equation}
where
\begin{equation}
T_{\mu\nu}({\bf k},\omega)\equiv\int d^3x' T_{\mu\nu}({\bf x'},\omega)e^{-i{\bf k}\cdot{\bf x'}}\,.
\label{T}
\end{equation}
Here we are supposing that  the radiation is observed in the wave zone, that is at distances $r\simeq|{\bf x}|$ much larger than the dimension $R=|{\bf x}'|_{max}$ of the source. 

The conservation equation for $T_{\mu\nu}({\bf x},t)$ is

\begin{equation}
\frac{\partial}{\partial x^\mu}T^{\mu}\,_{\nu}({\bf x},t)=0\,.
\end{equation}
Applying this result to the following equation
\begin{equation}
T_{\mu\nu}({\bf x},t)=T_{\mu\nu}({\bf x},\omega)e^{-i\omega t}+c.c.\,,
\end{equation}
gives
\begin{equation}
\frac{\partial}{\partial x^i}T^{i}\,_{\nu}({\bf x},\omega)-i\omega T^{0}\,_{\nu}({\bf x},\omega)=0\,.
\end{equation}
Multiplying with $e^{-i{\bf k}\cdot{\bf x}}$ and integrating over ${\bf x}$, we find that $T_{\mu\nu}({\bf k},\omega)$ is subject to the algebraic relations
\begin{equation}
k_{\mu}T^{\mu}\,,_{\nu}({\bf k},\omega)=0\,, 
\label{10.4.12}
\end{equation}
where $k_\mu$ is the vector ${\bf k}\simeq \omega {\hat x}$, $k^{0}\simeq \omega$. 

Now let us calculate the emitted power per unit solid angle along  a direction ${\bf{\hat{x}}}\simeq\frac{{\bf x}}{r}$ \cite{quadru}

\begin{equation}
\frac{dE}{d\Omega}=r^2 {\hat x}^i \left\langle T^{0i}\right\rangle\,.
\end{equation}
From Eq. (\ref{tensopol}), we obtain
\begin{equation}
\frac{dE}{d\Omega} = \frac{r^2({\bf k} \cdot {\hat x})k^0}{16\pi G}\left[e^{\lambda\nu*}({\bf x},\omega)e_{\lambda\nu}({\bf x},\omega)-\frac{1}{2}|e^{\lambda}_{\lambda}|^2\right]\,,
\end{equation}
 and after some algebra, we get
\begin{equation}
\frac{dE}{d\Omega} =\frac{G\omega^2}{\pi}\left[ T^{\lambda\nu*}({\bf k},\omega) T_{\lambda\nu}({\bf k},\omega)-\frac{1}{2} |T^{\lambda}\,_{\lambda}({\bf k},\omega)|^2\right]\,.
\end{equation}
The problem is  solved once we have calculated the Fourier transform (\ref{T})
\begin{equation}
\frac{dE}{d\Omega} = 2 G \int d \omega\;\omega^{2}[T^{\lambda \nu 
*}(\textbf{k},\omega) T_{\lambda \nu}(\textbf{k},\omega) - 
\frac{1}{2} |T^{\lambda}_{\lambda}(\textbf{k},\omega)|^{2}]\,.\label{10.4.13}
\end{equation}
It is convenient to express this result in terms of the purely space like components of $T^{\lambda\nu}({\bf k},\omega)$. From Eq. (\ref{10.4.12}),
we have
\begin{eqnarray}
T_{0i}({\bf k},\omega)&=&-{\hat k}^jT_{ji}({\bf k},\omega)\,,\nonumber\\
\nonumber\\
T_{00}({\bf k},\omega)&=&{\hat k}^i{\hat k}^jT_{ji}({\bf k}\,,\omega)\nonumber\\
\end{eqnarray}
where ${\bf {\hat k\simeq}}{\bf k}\ \omega\simeq {\hat x}$. Using these results in Eq.(\ref{10.4.13}) gives
\begin{equation}
\frac{dE}{d\Omega} = \frac{G\omega^2}{\pi} \Lambda_{ij, lm}({\hat k})T^{ij*}({\bf k},\omega)T^{lm}({\bf k},\omega)\,,
\label{10.4.14}\end{equation}
where the projection operator is \cite{gravitation}
\begin{eqnarray}
\Lambda_{ij, lm}({\hat k})&\simeq& \delta_{il}\delta_{jl}-2{\hat k}_j{\hat k}_m\delta_{il}+\frac{1}{2}{\hat k}_i{\hat k}_j{\hat k}_l{\hat k}_m+\nonumber\\
&&-\frac{1}{2}\delta_{il}\delta_{lm}+\frac{1}{2}\delta_{ij}{\hat k}_l{\hat k}_m+\frac{1}{2}\delta_{lm}{\hat k}_i{\hat k}_j\,.
\label{10.4.15}
\end{eqnarray}
If the energy-momentum tensor is a sum of individual Fourier components as 
\begin{eqnarray}
T_{\mu\nu}({\bf x},t)=\sum_\omega e^{i\omega t}T_{\mu\nu}({\bf x},\omega)+c.c
\end{eqnarray}
then the field $h_{\mu\nu}$ will look like a sum of  plane waves. The gravitational energy-momentum tensor is then  given by a double sum over these Fourier components, but all cross-terms drop out when we average over a time interval which is long if compared with the longest "beat period", that is, the reciprocal of the shortest frequency difference. The energy is thus given by a sum of terms like (\ref{10.4.14}), one for each frequency in the source \cite{weimberg}.
Suppose, on the other hand, that the energy-momentum tensor is a Fourier integral. Then $h_{\mu\nu}$ in the wave zone looks like an integral over $\omega$ of the individual plane waves, and the gravitational energy-momentum tensor is given by a double integral $\int \int d\omega d\omega'$ of products of these terms. The integrand, again, has time dependence $exp(-i(\omega-\omega')t)$, but now there is no longest beat period, so instead of computing the average power, we calculate the total emitted energy. This is given by integrating the energy over the whole time, and the effects is to replace the factors $e^{-i\omega t} e^{i\omega' t}$ in the double integral for the power with
\begin{eqnarray}
\int_{-\infty}^{\infty}exp(-i(\omega-\omega')t)dt=2\pi \delta(\omega-\omega')\,.
\end{eqnarray}
The energy per solid angle, emitted in a direction ${\hat{\bf k}}$ is thus a single integral:
\begin{eqnarray}
\frac{dE}{d\Omega}=2G\int_0^{\infty} \omega^2\left[^{\lambda \nu 
*}(\textbf{k},\omega) T_{\lambda \nu}(\textbf{k},\omega) - 
\frac{1}{2} |T^{\lambda}_{\lambda}(\textbf{k},\omega)|^{2}\right]d\omega\,,\nonumber\\
\label{a1}
\end{eqnarray}
For a free particle, we have ,
\begin{eqnarray}
T_{\mu \nu}(x) = m \int d \tau\; \dot{\xi}_{\mu}\;\dot{\xi}_{\nu} \;
\delta^{4}(x-\xi(\tau))\label{a3} \\
\xi^{\mu}(\tau) = (\gamma \tau, \gamma \textbf{v}\;\tau), 
 \;\;\;\;\;\dot{\xi}^{\mu}\dot{\xi}_{\mu} 
= -1.\nonumber\\\label{a4}
\end{eqnarray}
where $\gamma$ is the Lorentz factor.
This gives us the Fourier transform,
\begin{equation}
T_{\mu \nu} (\textbf{k},\omega) = m\;\dot{\xi}_{\mu}\dot{\xi}_{\nu} \;
\delta(\gamma \omega (1-\hat{\textbf{k}}\cdot \textbf{v})).\nonumber\\
\label{a5}
\end{equation}
The argument of this  $\delta$-function is the Lorentz invariant 
quantity $\dot{\xi}_{\mu}\;k^{\mu}$. It tells us that the radiation 
goes out along a cone, just as with familiar Cherenkov radiation; and 
this can only happen for velocities $v$ that are greater than $c=1$, in our units.
Now we put (\ref{a5}) into (\ref{a1}) and get,
\begin{equation}
\frac{dE}{d\Omega} = 2G \int d\omega \;\omega^{2}\; \frac{m^{2}}{2} 
\delta (\gamma \omega (1-\hat{\textbf{k}}\cdot 
\textbf{v}))\;\delta(0).\label{a6}
\end{equation}
It is worth noticing that
\begin{equation}
\delta(0) = \frac{1}{2\pi} \int d \tau e^{i (0)\tau} = \frac{\Delta 
\tau}{2\pi} = \frac{\Delta t}{2 \pi \gamma}.\label{a7}
\end{equation}
where $\Delta t$ is the time interval over which we observe this 
radiation process \cite{mpla}.

Now we integrate over all angles and get the rate of energy emission, that is 
\begin{equation}
\frac{\Delta E}{\Delta t} = G\;\frac{m^{2}}{\gamma^{2}\;v} \int 
d\omega\;\omega.\label{a8}
\end{equation}
We need to introduce some cut-off for the integral over $\omega$,  and 
to this end, let us consider the  basic quantum relation, $E = \hbar \omega$, 
where  the emitted quantum energy cannot be larger than 
the total energy $E$ of the particle.  Thus we get
\begin{equation}
\frac{\Delta E}{\Delta t} \sim \frac{G}{2}\;\frac{m^{2}}{\gamma^{2}\;v} 
\left(\frac{E}{\hbar}\right)^{2}.
\end{equation}
Noting that $E = m\gamma$, and re-introducing  the constant $c$, we get 
\begin{equation}
\frac{\Delta E}{\Delta t} \sim 
\frac{G}{2}\;\frac{m^{4}c^{4}}{\hbar^{2}\;v}.\label{a9}
\end{equation}

By substituting the values of the constants, we get the order of magnitude of the process, that is 
\begin{equation}
\frac{\Delta E}{\Delta t} \sim 
\left(\frac{m\,c^{2}}{eV}\right)^4\left(\frac{c}{v}\right)\times 10^{-41} \frac{eV}{sec}\,.\label{10}
\end{equation}

At this point, some consideration are  in order for Eq.(\ref{a9}).
 We find that the emitted energy  depends on  massless spin-$2$  (standard graviton),  
massive spin-$0$ and massive spin-$2$  modes.
Then according to the following values of the masses

\begin{equation}\label{energy_value}
\left\{\begin{array}{ll}
m_{spin2}\neq0\qquad \text{massive}\\\\
m_{spin2}=0\qquad \text{massless}\\\\
m_s\neq0\qquad \text{scalar}\\\
\end{array}\right.\end{equation}
we have different values for the energy.
As an example, we can consider, the upper bounds on the graviton mass, $m_g$, that come from direct or indirect observations of gravitational waves.  
A constraint on the graviton mass   comes from indirect evidence for the emission of gravitational waves from binary pulsars \cite{17}.  The upper limit is of the order
\begin{equation}
m_g\sim 7.6\times10^{-20}\,eV\,,
\label{masgpulsar}
\end{equation}
 Inserting  this value in Eq. (\ref{10}), we obtain  the emitted energy of the order
 
\begin{equation}
\frac{\Delta E}{\Delta t} \sim 
3.33\times 10^{-118} \left(\frac{eV}{sec}\right)\,.\label{pulgra}
\end{equation}
Furthermore, graviton masses of the order $m_g=10^{-30}eV$ could be detected by observing the characteristic signature of a  a strong monochromatic signal in gravitational wave detectors due to relic gravitons at a frequency which falls in the range for both of space based (LISA) and earth based (LIGO-VIRGO) gravitational antennas \cite{LVL}, in  the frequency interval  $10^{-4}Hz \leq f\leq10kHz$. In this case, we get

\begin{equation}
\frac{\Delta E}{\Delta t} \sim 
 10^{-161} \left(\frac{eV}{sec}\right)\,.\label{gravwav}
\end{equation}

On the other hand, considering a gravitational state of  mass order  $m_g=100\,GeV$,  that is a Higgs-like particle, we obtain 

\begin{equation}
\frac{\Delta E}{\Delta t} \sim 
10^{3} \left(\frac{eV}{sec}\right)\,.\label{mHiggs}
\end{equation}

and finally if $m_g = 1\,TeV$, we have

\begin{equation}
\frac{\Delta E}{\Delta t} \sim 
10^{7} \left(\frac{eV}{sec}\right)\,.\label{mLHC}
\end{equation}
In the last two cases, we are considering that massive gravitational  aggregates could emerge in high-energy experiments like those now running at LHC (CERN). See also \cite{basini} for details.

These results are a clear indication  that a gravitational Cherenkov radiation could be detected ranging from very low energy scales up to TeV scales as soon as further gravitational degrees of freedom are considered. A detailed experimental analysis will be reported in a forthcoming paper.


\begin{thebibliography}{99}


\bibitem{PRnostro}S. Capozziello, M. De Laurentis, {\it Physics  Reports } {\bf 509}, 167, (2011).

\bibitem{birrell}
N.D. Birrell and P. C. W.  Davies, {\it Quantum Fields in Curved Space}, Cambridge Univ. Press, Cambridge (1982).

\bibitem{basini}
S. Capozziello, G. Basini, M. De Laurentis,  {\it  The European Physical Journal C} {\bf 71}, 1679, (2011).

\bibitem{greci}
C. Bogdanos, S. Capozziello, M. De Laurentis, S. Nesseris, {\it Astroparticle Physics} {\bf 34}, 236, (2010).


\bibitem{felix} 
S. Capozziello, M. De Laurentis, C. Corda, {\it  Physics Letter B} {\bf 699}, 255, (2008).

\bibitem{LHC}http://lhc.web.cern.ch/lhc/

\bibitem{francaviglia1}
S. Nojiri, S.D. Odintsov, {\it Physics Reports} {\bf 505}, 59 (2011).
\bibitem{francaviglia2}
S. Capozziello, M. De Laurentis, M. Francaviglia, {\it Astrop. Phys.} {\bf 29} 125, (2008).
\bibitem{francaviglia3}
S. Nojiri, S.D. Odintsov, {\it Int. J. Geom. Meth. Mod. Phys.} 4, 115, (2007).
\bibitem{francaviglia4}
S. Capozziello,  M. Francaviglia, {\it Gen. Rel. Grav.}   \textbf{40},357, (2008).
\bibitem{francaviglia5}
S. Capozziello, M. De Laurentis, V. Faraoni, {\it The Open Astr. Jour} , \textbf{2}1874, (2009).

\bibitem{gupta}
A. Gupta, S.  Mohanty,  M. K.  Samal, {\it Class. Quant. Grav.} {\bf 16}, 291 (1999).

\bibitem{gaetano}
G. Lambiase, {\it Europhys. Lett.} {\bf 56}, 778 (2001).

\bibitem{mpla}
C. Schwartz, {\it Mod. Phys. Lett.} {\bf A}, {\bf 26}, 2223 (2011).

\bibitem{Nunez:2004ts}
  A.~Nunez and S.~Solganik,
 {\it Phys.\ Lett.} B {\bf 608}, 189 (2005).
\bibitem{Chiba:2005nz}
  T.~Chiba, {\it JCAP} {\bf 0503}, 008 (2005).
\bibitem{Stelle:1977ry}
  K.~S.~Stelle,
  {\it Gen.\ Rel.\ Grav.} {\bf 9}, 353 (1978).
  
  \bibitem{vanDam:1970vg}
  H.~van Dam and M.~J.~G.~Veltman,
  {\it Nucl.\ Phys.} B {\bf 22}, 397 (1970).
 \bibitem{gravitation}C. W. Misner, K. S. Thorne
and J. A. Wheeler,  {\it Gravitation},  W.H.Feeman $\&$ Co.,  New York
(1973).
\bibitem{quadru}M. De Laurentis, S. Capozziello, {\it Astroparticle Physics}, {\bf35},  257 (2011).

\bibitem{weimberg}S. Weinberg, {\it Gravitation and Cosmology: Principles and Applications
of the General Theory of Relativity}, John Wiley \& Sons, New
York (1972).

\bibitem{17}J. H. Taylor, {\it Rev. Mod. Phys.} 66, 711 (1994).

\bibitem{LVL}http://www.ligo.org/pdfpublic/camp.pdf; http://www.ligo.org/pdfpublic/hou
gh02.pdf\\
http://www.virgo.infn.it
\\ http://www.lisa.nasa.gov; http://www.lisa.esa.int

\end{thebibliography}
\end{document}